# Fully printed, all-carbon, recyclable electronics

**Authors:** Nicholas X. Williams[1], George Bullard[2], Nathaniel Brooke,[1] Michael J Therien[2], Aaron D. Franklin[1,2]*

**Affiliations:**

[1]Department of Electrical and Computer Engineering, Duke University, Durham NC 27708, USA

[2]Department of Chemistry, Duke University, Durham NC 27708, USA

*Correspondence to: aaron.franklin@duke.edu, TEL: +1-919-681-9471

**Abstract:** The rapid growth of electronic waste must be curtailed to prevent accumulation of environmentally and biologically toxic materials, which are essential to traditional electronics[1]. The recent proliferation of transient electronics has focused predominantly on biocompatibility[2,3], and studies reporting material recapture have only demonstrated reuse of conducting materials[4–6]. Meanwhile, the ideal solution to the electronic waste epidemic — recapture and reuse of all materials — has been largely neglected. Here we show complete recyclability of all materials in printed, all-carbon electronics using paper substrates, semiconducting carbon nanotubes, conducting graphene, and insulating crystalline nanocellulose. The addition of mobile ions to the dielectric produced significant improvements in switching speed, subthreshold swing, and among the highest on-current for printed transistors. These devices evinced superlative stability over 6 months, after which they are shown to be controllably decomposed for complete recycling of materials and re-printing of devices with similar performance to baseline devices. The printing of all-carbon, recyclable electronics presents a new path toward green electronics with potential to mitigate the environmental impact of electronic waste. We anticipate all-carbon, recyclable electronics to be a watershed, facilitating internet-of-everything applications, such as ubiquitous sensors for continuous monitoring of diseases or environmental conditions, while preserving carbon neutrality in the device lifecycle.

**Main Text:** Electronic waste (e-waste) is an escalating global concern, contributing high levels of often toxic pollutants that are one of the fastest growing waste streams worldwide[7]. One promising method to decrease the ecological and health impacts of e-waste is physically transient electronics — devices designed for controlled disintegration after use[3]. Yet this approach, which has been largely used in biologically integrated sensors[2] and therapeutics[8], still results in accumulation of silicon-based materials and carbon-nanostructured waste, which are themselves undesirable e-waste[9]. Fully recyclable electronics remain the goal in order to eliminate or recapture and reuse environmentally and biologically damaging components of e-waste.

Previous efforts to realize recyclable electronics have focused on recapture of a liquid metal conductive trace[4] or the degradation of a single component[10]. For significant global impact, a unified system must be considered, combining degradation and reuse of all components. Further, recyclable electronics can still have negative impacts because their high-temperature manufacture requires both significant energy and environmentally damaging processing conditions[11]. The environmental costs of manufacturing electrical devices can be substantially reduced by printing at room temperature[12]. One promising approach is the utilization of all-carbon components[13], as semiconducting carbon nanotubes (CNTs) and conducting graphene both have long histories in printed electronics[14,15], while cellulose has been used previously both as substrate[16] and dielectric[17]. Yet, there have been no complete demonstrations of all-carbon printed electronics, largely due to an absence of solution-processable carbon dielectrics, with previous demonstrations incorporating inorganic components[18–20]. Despite the use of cellulose paper as a dielectric in high-power transformers[21], crystalline nanocellulose (CNC) has been largely overlooked as a standalone printable dielectric and relegated to use as a binder or, at most, in tandem with a high-k dielectric[22,23].

Here we report all-carbon recyclable electronics (ACRE), fully printed on paper and made possible by developing a CNC dielectric ink printable at room temperature with compatible CNT and graphene inks. CNC dielectric performance was significantly improved through the addition of 0.15 mM NaCl to the aqueous ink, allowing for higher transistor switching speeds with low subthreshold swing and high on-current. We accomplished this using all-carbon components, a cellulose substrate, room-temperature direct-write aerosol jet printing (AJP), and, most uniquely, crystalline nanocellulose as the dielectric. Furthermore, we show the complete recyclability of these printed electronic devices, including the recapture and reuse of the constituent nanomaterials.

We investigated the use of crystalline nanocellulose (Fig. 1a) as a standalone printed dielectric. Cellulose is the most abundant biopolymer on earth, making it relatively inexpensive, while also being completely biodegradable[22]. Testing the relationship between dynamic viscosity and CNC concentration, we found that all tested concentrations of CNC in water displayed shear-thinning behavior with viscosities ranging across two orders of magnitude. Furthermore, all inks maintained viscosities orders of magnitude greater than water at low sheer rates (Fig. 1b). Given this result, AJP is the ideal direct-write deposition method as it allows for the printing of high-viscosity fluids compared to inkjet printing[24]. AJP functions via the ultrasonication of ink to create a suspension of ink constituents in 1-5 $\mu$m droplets (Fig. 1c,d). At low concentrations and flow rates, deposited microdroplets evaporate too rapidly to realize a coherent film and form a porous structure (Fig. 1e,f), whereas at sufficiently high flow rate and concentration monolithic films are formed that are ideal for insulating electrical charge. With very few exceptions[25], printable dielectrics require elevated post-processing temperatures to exhibit desirable insulating behavior[26], which is undesirable because extended processing increases cost and environmental impact. To minimize environmental footprint, CNC films were printed at room temperature without observable performance degradation (Extended Data Fig. 1).

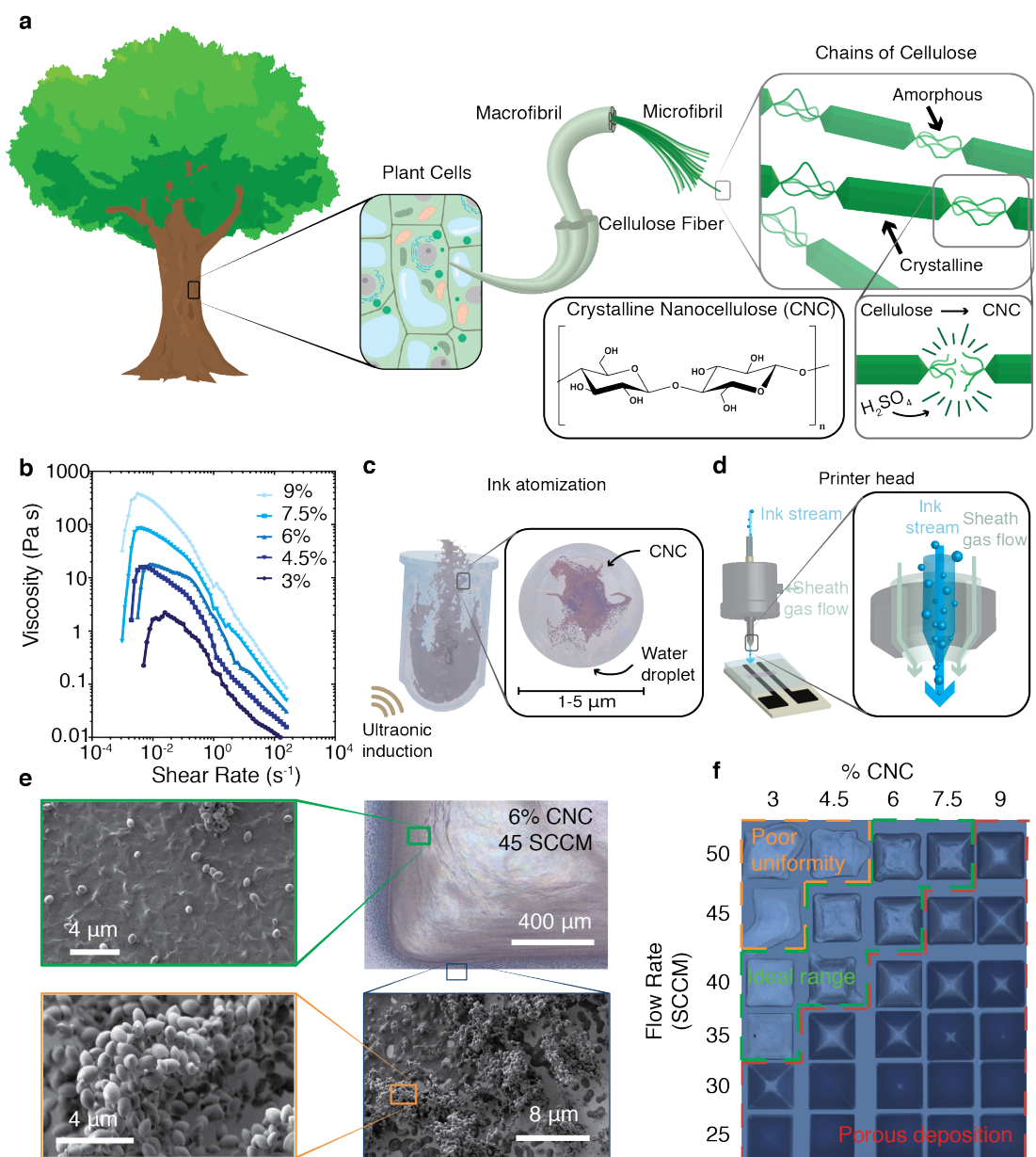

Fig. 1. **Nanocellulose printing.** (a) Schematic of hierarchical structure of crystalline nanocellulose. (b) Dynamic viscosity plotted versus sheer rate for a range of CNC concentrations. (c) Schematic of AJP ultrasonication process with magnified view of ink droplet. (d) Schematic of printer head with magnified view of printer nozzle. (e) Scanning electron microscope (SEM) images of a printed CNC film at various magnifications. (f) Grid of printed CNC squares as a function of flow rate (measured in SCCM) and CNC concentration.

Using this CNC dielectric ink, all-carbon thin-film transistors (TFTs) were printed on a paper substrate with graphene source/drain/gate electrodes and a CNT channel (Fig. 2a,b). Previous reports claiming "all-carbon" electronics actually incorporate inorganic components, focusing only on a carbonaceous channel and electrodes[18–20]; the printed CNC gate dielectric in our ACRE-TFTs yields a fully all-carbon device. The CNC films exhibited frequency-dependent dielectric behavior consistent with ion-gel dielectrics, facilitated by absorbed water, which mobilized ions in the film to yield a double-layer capacitance (Extended Data Fig. 2,3)[27]. Ion gels generally provide increased freedom in transistor gate placement, leading this part of the design to often be overlooked in favor of simple side-gated arrangements to reduce fabrication complexity. However, for these all-carbon TFTs, we discovered a gate position-dependent modulation of the channel charge, as shown in the surface map of the transistor's on-current/off-current ratio ($I_{on}/I_{off}$) versus gate probe tip position (Fig. 2c-d). Unlike ion-gel dielectrics, almost no switching behavior was observed far afield from the channel, hence a top-gated structure was critical in these devices.

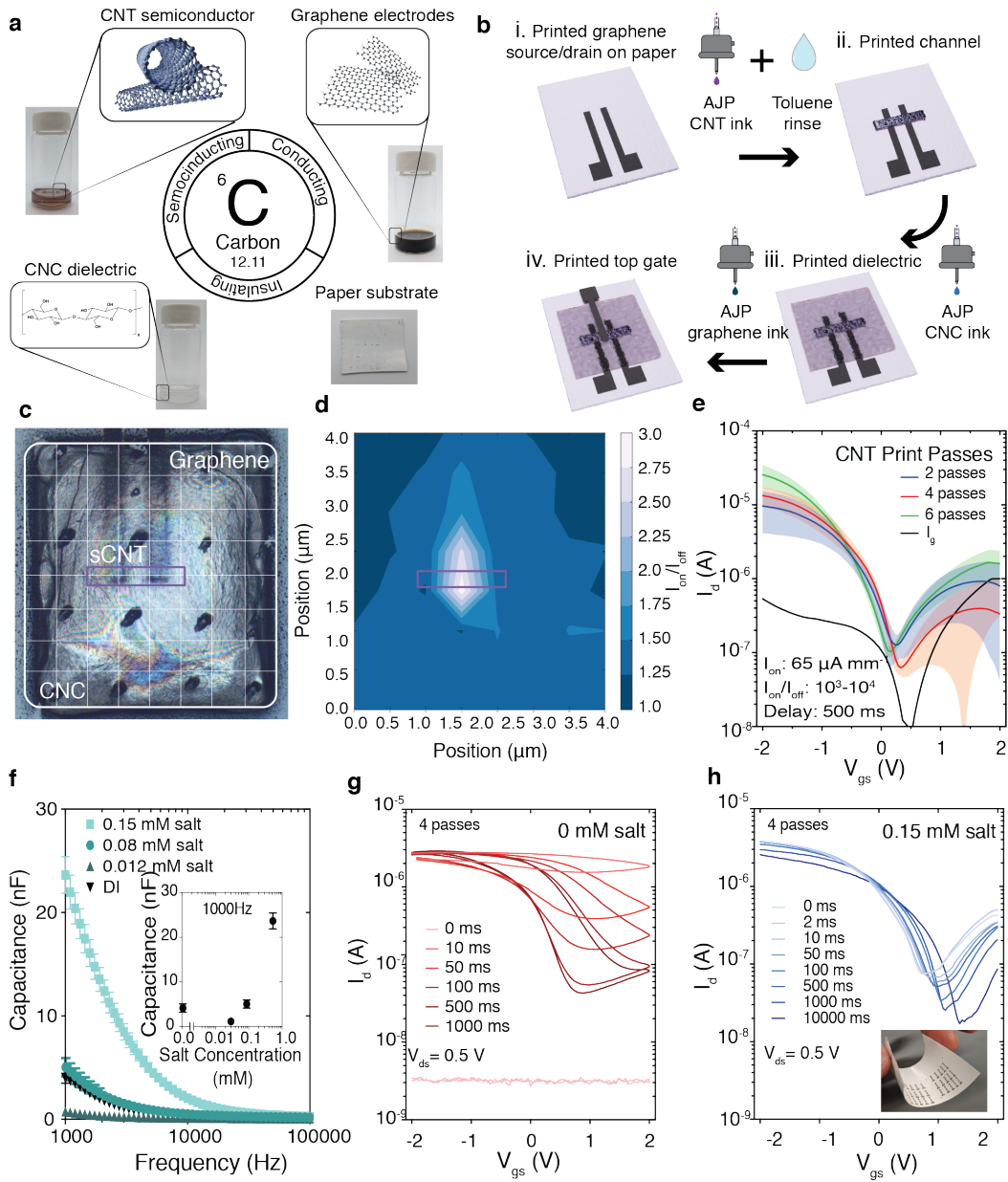

Fig. 2. **CNC-based all-carbon TFT printing and testing.** (a) All-carbon electronic components. (b) Printing fabrication schematic. (c) Optical image of all-carbon TFT used for (d) surface map of transistor $I_{on}/I_{off}$ as a function of gating location. (e) Subthreshold curves as a function of CNT print passes. Data represent average (line) ± standard deviation (shaded region) of 12 devices without added salt at a $V_{ds}$ of -0.5 V. (f) Frequency-dependent capacitance vs ink salt concentrations with values at 1000 Hz inset. Subthreshold curves measured with different delay times for devices with (g) no added salt and (h) addition of 0.15 mM NaCl to CNC ink; inset is photo of an array of all-carbon transistors. All devices in this report have a channel length of 250 μm and a channel width of 200 μm.

The impact of CNT film density was investigated for the all-carbon TFTs with 2, 4, and 6 printing passes of CNT ink; all processing was performed at room temperature except for a 10-minute toluene soak at 80 °C after nanotube printing. While previous reports indicate promising performance with a lower density of AJP CNTs[28], 4 passes of CNT ink provided transistors with optimal performance, demonstrating an $I_{on}/I_{off}$ of $10^3$-$10^4$, an $I_{on}$ of $65.00\pm18.25$ $\mu$A mm$^{-1}$, and a subthreshold swing (SS) of $132\pm70$ mV dec$^{-1}$ (Extended Data Fig. 4). Tolerance to increased CNT density is attributed to the higher photo paper roughness compared to a silicon surface. Even with the increased roughness, and thus reduced deposition uniformity, these all-carbon devices rival the overall highest performance printed transistors (from any material) and are among the best performing transistors on paper, particularly in terms of on-current at low voltages (~150 $\mu$A/mm at 1 V) (Extended Data Fig. 5 and Table S1).

Since performance of ionic-based dielectrics depends on the diffusion of ions to establish a double layer of charge (Extended Data Fig. 6), switching speed is traditionally slower than solid-state dielectrics[29]. This is observed by the frequency-dependent capacitance (Fig. 2f) and the transistor switching dependence on delay time (Fig. 2g) – the duration that each gate voltage value is sustained before the initiation of current measurement. Almost no modulation is observed at delay times of 10 ms ($I_{on}/I_{off}$ of 1.5), whereas the transistor behaves as expected at a 500 ms delay time, with an $I_{on}/I_{off}$ of 66.1.

This frequency-dependent behavior indicates limited ionic mobility, which is ultimately determined by the ion source in the dielectric. Nanocellulose is produced by acid hydrolysis of cellulose, a process that sulfonates the cellulose, leaving it charged[30]. To confirm this, the addition of 0.012 mM salt to the ink solution (Extended Data Fig. 7) was shown to screen the CNC surface charge and nearly eliminate the frequency-dependent capacitance (Fig. 2f and insert). Given that the surface charge is bonded to the CNC backbone, ion mobility is minimal, and the device requires extended delay times to operate. Once this surface charge is screened, any addition of salt ions function as highly mobile carriers, improving device responsiveness. The addition of 0.15 mM NaCl almost entirely eliminates the dependence on delay time (Fig. 2h). While device performance is still dependent on ion mobility, only a 21% decrease in $I_{on}/I_{off}$ is observed when measurement time is decreased from 500 ms to 10 ms as opposed to a 98% decrease without the addition of the mobile ions.

The all-carbon inks allow for the development of printed all-carbon recyclable electronics (Fig. 3a). The ACRE transistors exhibit phenomenal stability when stored in air over the course of 6 months (Fig.

3b). Once their utility is exhausted, the components can be broken down and the inks recycled for later reuse. Previous reports show significant performance alterations upon recycling[6] or recycle only the conductive trace[4]. Ultrasonic recycling of ACRE-TFTs produced inks nearly indistinguishable from the original CNT ink (Fig. 3c-e) with functional, although less conductive graphene ink (Fig. 3f). Furthermore, we reprinted the recycled CNTs into transistors with nearly identical performance to the TFTs from new ink (Fig. 3f); the slight decrease in on-current is attributed to the decreased printed CNT density (Extended Data Fig. 8) and is commensurate with batch-to-batch variations. Fully recycled ACRE-TFTs showed only minimal decrease in on-current, attributed to the lower conductivity of the recycled graphene that can be improved through reformulation of the recycled ink. Many physically transient devices degrade over the course of use[3]; yet, these ACRE devices exhibited stable performance over 6 months and are capable of controlled recycling upon completion of utility, such as use in biosensing. To demonstrate the applied sensing utility of the ACRE platform, a fully printed, paper-based biosensor from an ACRE-TFT for lactate sensing was successfully realized and shown to yield an expanded measurement range (compared to a simple two-electrode device), reaching concentrations relevant for medical diagnosis of sepsis (>2 mM) (Extended Data Fig. 9).

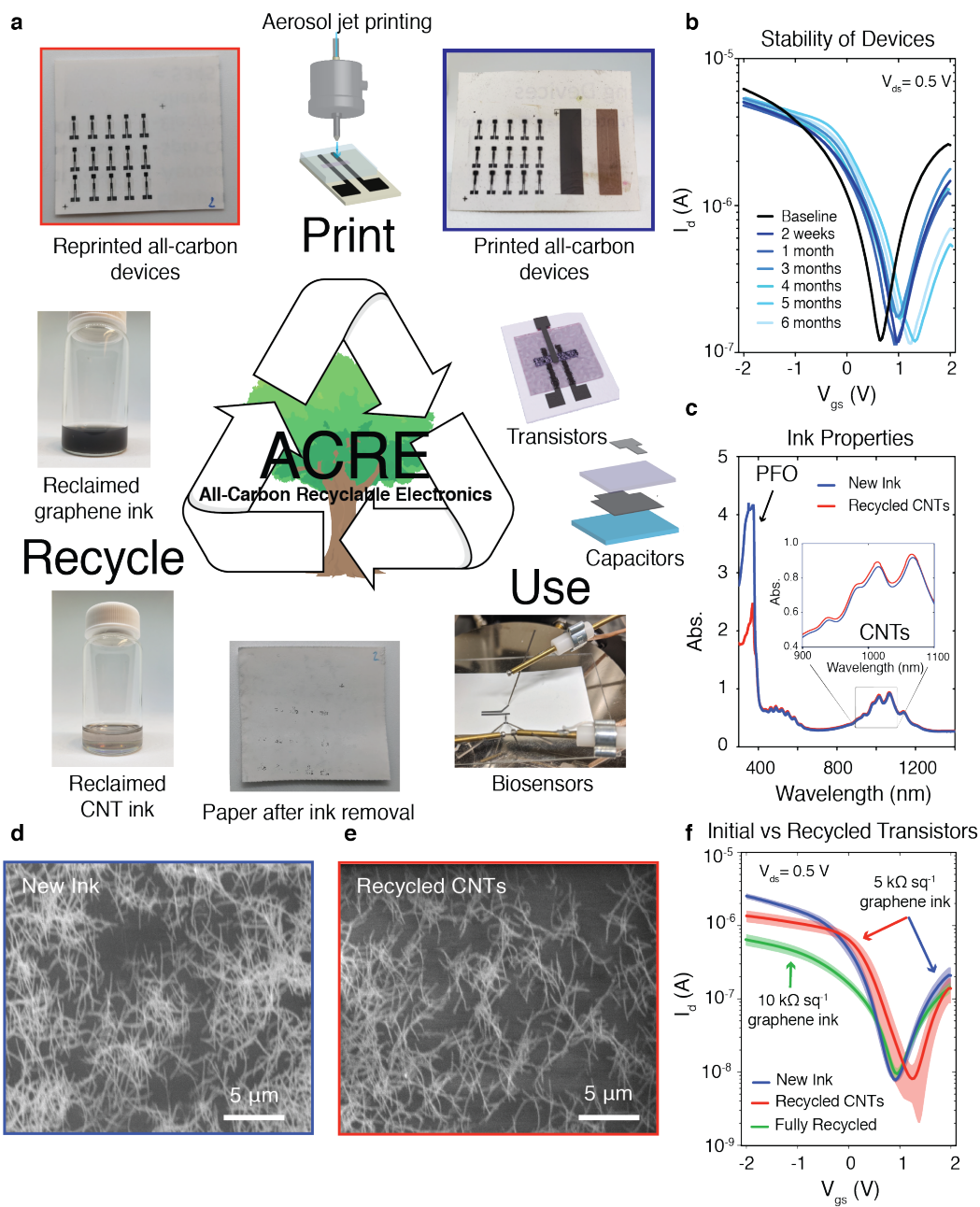

Fig. 3. **ACRE-TFTs with controlled recycling demonstrated.** (a) Schematic of ACRE system demonstrating printing, use (see Extended Data Fig. 9 for biosensing demonstration), recycling, and then reprinting, reuse, etc. Top right insert shows array of ACRE-TFTs with excess graphene (left) and CNTs (right) printed to the side. (b) Stable TFT characteristics over 6 months storage in air. Transistor fabricated without salt and run with 500 ms delay time. (c) UV-vis-NIR spectra of new and recycled CNT inks. SEM images of printed (d) new and (e) recycled CNT inks. (f) Subthreshold curves of transistors from new and recycled inks. Data represent average ± standard deviation of 4 devices with a salt concentration of 0.15 mM in the CNC and a delay time of 10 ms.

The establishment of a frequency-independent, printable CNC dielectric ink has enabled the demonstration of fully printed and recyclable all-carbon transistors on paper. Printing reduces processing energy requirements and the all-carbon components yield devices that are air-stable with exceptional performance. Altogether, this ACRE platform offers an environmentally friendly approach to printing custom electronic devices for the expanding internet-of-everything, from transistors to biosensors.

**Materials**

Crystalline nanocellulose (CNC-Slurry) was purchased from Cellulose Lab at a concentration of 12% solids. To prepare the ink for printing, CNC was diluted in water to concentrations ranging from 3-9% W/W CNC. Semiconducting carbon nanotubes (CNTs) (IsoSol-S100® Polymer-Wrapped Nanotubes) were purchased from NanoIntegris. To prepare the ink for printing, the vial of as-purchased ink was ultrasonicated for 30 minutes to re-disperse the CNTs. The ink was diluted to a concentration of 0.05 mg ml$^{-1}$ in toluene for printing. Graphene ink (805556-10ML) was purchased from Sigma-Aldrich. The ink was diluted to 1/3 the initial concentration in water for printing. Paper (Gloss White) was purchased from FedEx and used as is. Phosphate buffered saline (1xPBS) was purchased from Sigma-Aldrich and used as is. Lactate oxidase (LOx) powder (L9795) was purchased from Sigma-Aldrich. The powder was mixed with 1xPBS to achieve a concentration of 200 U ml$^{-1}$ and vortexed for 5 minutes to dissolve. 50 $\mu$L of the protein was distributed into Eppendorf flasks, and flash-frozen in liquid nitrogen. The protein was thawed by placing one Eppendorf flask in an ice water bath, as a slower thaw reduces the risk of LOx denaturization. Sodium L-lactate was purchased from Signa-Aldrich and concentrated to 10 mM in 1xPBS. Silver nanowires (AgNW) of length ~4 $\mu$m were synthesized using the method described in a previous publication[31], and suspended in deionized (DI) water at a concentration of 10 mg ml$^{-1}$.

**Methods**
***Aerosol jet printing***: All printing was performed on an AJ-300 printer (Optomec; Albuquerque, NM) with the platen temperature at room temperature and the ink held at a constant 20 °C throughout printing.

- ***Graphene printing***: A 150 $\mu$m diameter nozzle was used to print all graphene traces and films with a print speed of 2 mm s$^{-1}$. A sheath flow rate of 25 standard cubic centimeters per minute (SCCM), an atomizer flow rate of 40-50 SCCM, and an ultrasonic inducer current of 340-350 mA was used. One pass of graphene ink was used in all prints with new ink and 4 print passes were used for recycled graphene ink due to modifications such as concentration and polymer fraction that resulted in higher resistance of the resulting ink (note: these should be addressable through reformulation of the recycled ink to match that of the new ink). As soon as a coherent line was observed, printing was initiated.

- **Semiconducting carbon nanotube printing:** A 150 $\mu$m diameter nozzle was used to print all CNT films with a print speed of 2 mm s$^{-1}$. A sheath flow rate of 40 SCCM, an atomizer flow rate of 23 SCCM, and an ultrasonic inducer current of 340-350 mA was used. 2-6 passes of CNT ink were used for all prints. The ink temperature was allowed to equilibrate in the printer under ultrasonication for 30 minutes before the atomizer flow was initiated. CNT flow was allowed to stabilize for 10 minutes before printing.

- **Crystalline Nanocellulose Printing:** A 300 $\mu$m diameter nozzle was used to print all CNC films with a print speed of 5 mm s$^{-1}$. A sheath flow rate of 23 SCCM, an atomizer flow rate of 45 SCCM, and an ultrasonic inducer current of 340-350 mA was used. 1 pass of CNC ink was used for all prints. As soon as a coherent line, without excessive overspray was observed, printing was initiated.

- **Silver Nanowire Printing:** A 150 $\mu$m diameter nozzle was used to print the AgNW films with a print speed of 1 mm s$^{-1}$. A sheath flow rate of 23 SCCM, an atomizer flow rate of 45 SCCM, and an ultrasonic inducer current of 340-350 mA was used. 3 total passes were used to enhance conductivity.

**_Capacitor fabrication_**: Capacitors were fabricated on glass substrates. To prepare the substrate, the glass slide was first ultrasonicated in isopropyl alcohol (IPA) for 10 minutes, rinsed with deionized (DI) water, and dried with nitrogen ($N_2$), then ultrasonicated in acetone for 10 minutes, rinsed with DI water, and dried with $N_2$. 30 nm of gold was evaporated using a CHA Industries Solution E-Beam evaporator for the bottom electrode. Then CNC was printed as the dielectric using the above parameters. Finally, AgNWs were printed as the top electrode using the above parameters.

**_Transistor fabrication_:** To print the all-carbon transistors, a piece of photo paper was placed onto the aerosol jet printer platen. First, source and drain electrodes were printed using graphene ink using the above parameters, then without removal of the substrate from the platen, the CNT channel was printed using the above parameters. After CNT printing, the substrate was soaked in a toluene bath for 10 minutes at 80 °C to remove excess polydioctylfluorene (PFO). This process is required to achieve a sufficiently conductive transistor channel; note that other reports have demonstrated this polymer-removal process as also being feasible using a toluene rinse without removing the substrate from the printer platen, so long as the platen is held at an elevated temperature (80 ºC). The substrate was then placed back on the printer platen and the CNC ionic dielectric was printed using the above parameters. Finally, without further removal of the substrate from the printer platen, a graphene gate was printed atop the CNC. All devices in this report have a channel length of 250 $\mu$m and a channel width of 200 $\mu$m.

**_Lactate sensor fabrication_:** To print the all-carbon lactate sensor, a piece of photo paper was placed onto the aerosol jet printer platen. For the 2-electrode sensor, graphene working and common electrodes were printed using the above parameters. For the transistor-based sensor, first, source and drain electrodes were printed using graphene ink using the above parameters, then without removal of the substrate from the platen, the CNT channel was printed using the above parameters. After CNT printing, the substrate was soaked in a toluene bath for 10 minutes at 80 °C. The substrate was then placed back on the printer platen and the CNC dielectric was printed using the above parameters. Finally, without further removal of the substrate from the printer platen, a graphene working and common electrode were printed to the side of the device. Finally, a gate that is connected to the working electrode was printed on top of the channel.

**_Lactate Concentration Testing_:** All testing was performed on a Signatone H150W manual analytical probe station connected to an Agilent (Keysight Technologies) B-1500 Semiconductor Parameter Analyzer. For each test, 100 $\mu$L of 1xPBS was pipetted onto the paper substrate directly above the working and common electrodes. 1$\mu$L of LOx was pipetted into the PBS droplet and mixed 10 times with a pipette. A baseline measurement was taken. For the 2-electrode measurement – the voltage across the working and common electrode was swept from -1 to 1 V and then back from 1 to -1 V. For the transistor

measurement – the gate electrode was attached to the common electrode while the working electrode that was connected to the gate was allowed to float (Extended Data Fig. 9). During testing, the gate voltage was held at -0.5V and the source-drain voltage was swept from -2 to 2 V. Varying concentrations of lactate were dosed into the PBS-LOx solution and mixed 10 times with a pipette. After lactate dosing and mixing, the solution was allowed to stabilize for 2 minutes before a voltage sweep was initiated for both the 2-electrode and the transistor-based sensors.

***Capacitor heat testing***: A baseline frequency sweep was performed on a Signatone H150W manual analytical probe station connected to an Agilent (Keysight Technologies) B-1500 Semiconductor Parameter Analyzer. Then the capacitors were heated to a temperature of 80 °C on a hot plate for 1 hour. After an hour, the capacitor was removed from the hot plate and a frequency-capacitance sweep was performed every 30 seconds for 30 minutes.

***Capacitor vacuum testing***: Vacuum testing was performed using a cryogenic probe station (Lakeshore CRX-6.5K) in tandem with an Agilent (Keysight Technologies) B-1500 Semiconductor Parameter Analyzer. The capacitor was placed into the chamber and placed under vacuum of $10^{-5}$ Torr. The system was allowed 10 minutes to achieve a vacuum. At this point, the test was initiated. A frequency-capacitance sweep was performed every 30 seconds for the duration of the testing. After 12 minutes, the vacuum was purged and the top to the vacuum chamber was removed, exposing the capacitor to air. After 12 minutes in air, the top was replaced, and the system was pumped down to a vacuum. Then, to test whether the increase in capacitance was due to pressure or atmosphere, the system was brought up to atmospheric pressure in nitrogen and held for 12 minutes, following by a 12-minute hold in air with the vacuum chamber to removed. This was followed by a pump down, purge, and pump down cycle, each 12 minutes long.

***CNT printed density determination***: New and recycled inks were diluted to the same concentration (0.05 mg ml$^{-1}$ assuming that 100% of all deposited ink was recaptured) and aerosol jet printed onto a silicon wafer with 300 nm thick thermally grown silicon dioxide insulator. The printing parameters for each were identical with an atomizer flow rate of 40 SCCM, a sheath flow rate of 23 SCCM, an ultrasonic inducer current of 340-350 mA, and a printer speed of 2 mm s$^{-1}$. As with all CNT printing, the ink temperature was held at 20 °C to increase printing stability and each vial was allowed to reach thermal equilibrium with the ultrasonic bath for 30 minutes before the atomizer flow was initiated. An additional 10 minutes settling time ensured that the print had stabilized after the atomizer and sheath flows were activated. Given that the surface roughness of silicon is substantially lower than that of paper, only one pass of CNTs were used for characterization (Extended Data Fig. 8a,b). A SEM was used to take images of each print and the image processing software ImageJ was used to determine CNT density.

Once the images were loaded into the image processing software, the image was set to Binary (Extended Data Fig. 8c,d). An additional image using color thresholds is displayed in Extended Data Fig. 8e,f. Once this procedure was finalized, the area fraction was measured using the built-in ImageJ procedure. The ~10% decrease in printed CNT density provides some explanation for the slightly lower on-current from the recycled ink. This decrease in concentration may be due to slight error in measurement of ink volume usage during printing, error in ink dilution after recycling, incomplete removal of the CNTs from the substrate, or some combination the above.

***Instrumentation and Characterization***: SEM images were taken at the Shared Materials Instrumentation Facility (SMIF) at Duke University. Vacuum testing was performed using a Cryogenic Probe Station (Lakeshore CRX-6.5K) in tandem with an Agilent (Keysight Technologies) B-1500 Semiconductor Parameter Analyzer. Viscosity was measured using an AR-G2 Magnetic bearing rheometer.

***CNT recovery and purification***:

- ***Recovery procedure for spectra analysis***: The total volume of ink deposited was measured and this volume was used as a guide to determine total CNTs printed and the volume to dilute the recovered ink to post recycling. CNTs printed on gloss white paper were bath sonicated in 5 mL of toluene. After removal of the paper, the resulting mixture was concentrated by centrifuging at 4000 g for 15 minutes in an Amicon Ultracel 10k filter. A minimal amount of toluene was used to collect the CNTs from the filter and transferred to a clean vial. Toluene was added to match the initial printing volume (in order to achieve similar ink concentrations before and after recycling) before batch sonicating for 15 minutes. Electronic absorption spectra were collected using a 2 mm path quartz cell using a Varian 5000 spectrometer to evaluate concentration.

- ***Recovery procedure recycled CNT-TFT***: All-carbon transistors printed on gloss white paper were tip horn sonicated for 15 minutes (20 kHz; 12 W total power) at 0 ºC in toluene. After removal of the paper, the resulting mixture was centrifuged at 4000g for 15 minutes in an Amicon Ultracel 50k filter. This step was used to recover and concentrate the printed CNTs while removing other materials. A minimal amount of toluene was used to collect the CNTs from the filter which were subsequently transferred to a clean vial. Toluene was added to match the initial printing volume before batch sonicating for 15 minutes.

- ***Recovery procedure for fully recycled TFTs***: All-carbon transistors printed on gloss white paper were bath sonicated in 5 mL of toluene to remove CNTs. The paper was removed and rinsed with 1 mL of toluene and then allowed to dry. The resulting CNT mixture was concentrated by centrifuging at 4000 g for 15 minutes in an Amicon Ultracel 10k filter. A minimal amount of toluene was used to collect the CNTs from the filter and transferred to a clean vial. Toluene was added to match the initial printing volume before batch sonicating for 15 minutes. The dry gloss white paper was then subjected to bath sonication in 5 mL of water to remove the graphene electrodes. The paper was removed and rinsed with an additional 1 mL of water. The resulting mixture was passed through a cotton filter and then concentrated in vacuo. The recovered graphene was transferred to a clean vial and diluted with water to match the initial printing volume and then bath sonicated for 15 minutes.

**_Mobile ion calculation_**: Nanocellulose is formed by acid hydrolysis, which is a process where amorphous regions of cellulose are chemically hydrolyzed by sulfuric acid ($H_2SO_4$), leaving only the nanoscale crystalline region[32]. Recent reports indicate that the hydrolysis reaction used to synthesize nanocellulose from cellulose leaves the nanocellulose with a sulfate surface charge that is dependent upon the hydrolysis conditions[33]. We hypothesize that this, along with a very small addition of dissolved ions in the DI water used in the acid hydrolysis, is the intrinsic ion source that facilitates the ionic dielectric behavior in the CNC films. To test this hypothesis, inks consisting of desulfonated CNC (CNC solutions derived from removal of the sulfate surface charges) along with inks prepared utilizing ultrapure water were fabricated, and the capacitances of these samples tested (Extended Data Fig. 3). Both the desulfonation of the CNC and the removal of the small concentration of ions in the DI water decreased the resulting capacitance, and with the removal of both, nearly all frequency dependence was eliminated, leaving a poor dielectric.

To further understand the correlation between ionic charge and capacitance, mobile ions at varying concentrations were added to the CNC ink prior to deposition to screen the intrinsic charge. The addition of a small concentration of mobile ions (less than the remnant charge concentration from the synthesis), would decrease the observed capacitance because the added ionic charge screens the sulfate charge on the CNC, negating the charge in the system and eliminating the ionic dielectric behavior[34]. After the sulfate charge is overwhelmed by the added ionic charge, a significant increase in the dielectric capacitance should be observed. To test this hypothesis, first the concentration of sulfate charge must be calculated.

A recent report determined that the sulfonation of the CNC during acid hydrolysis produced a sulfur concentration, a direct corollary to surface charge, between 3-10 mg g$^{-1}$ CNC[30]. In this work, CNC was concentrated to 6% w/w for purposes of printability, hence sulfur concentration would be 180-600 mg g$^{-1}$, which equates to an average charge molarity of 0.012 mM. Thus to completely screen the CNC surface charge, 0.012 mM NaCl must be added to the 6% w/w CNC in DI $H_2O$ solution and all further additions of salt to the aqueous ink would increase the total mobile charge in the solution.

**Acknowledgments:** This work was performed in part at the Duke University Shared Materials Instrumentation Facility (SMIF), a member of the North Carolina Research Triangle Nanotechnology Network (RTNN), which is supported by the National Science Foundation (Grant ECCS-1542015) as part of the National Nanotechnology Coordinated Infrastructure (NNCI).

**Funding:** This work was supported by the Department of Defense Congressionally Directed Medical Research Program (CDMRP) under award number W81XWH-17-2-0045 and by the National Institutes of Health (NIH) under award number 1R01HL146849. G.B. and M.J.T. are grateful to the Air Force Office of Scientific Research for research support under award number FA9550-18-1-0222.

**Author contributions:** N.X.W. and A.D.F. conceived the idea for the project, designed the experiments, and analyzed the data. N.X.W. performed ink development, printing, device design, and characterization experiments. G.B. and M.J.T. performed and analyzed recycling experiments. N.B. performed and analyzed lactate sensor experiments. N.X.W. and A.D.F. wrote the manuscript. All authors discussed the results and commented on the manuscript.

**Competing interests:** Authors declare no competing interests.

**Data and materials availability:** All data needed to evaluate the conclusions herein are present in the paper or supplementary materials. The materials that support the findings of this study are available from the corresponding author upon reasonable request.

# Extended Data for

## Fully printed, all-carbon, recyclable electronics

Nicholas X. Williams[1], George Bullard[2], Nathaniel Brooke,[1] Michael J Therien[2], Aaron D. Franklin[1,2]

Correspondence to: aaron.franklin@duke.edu

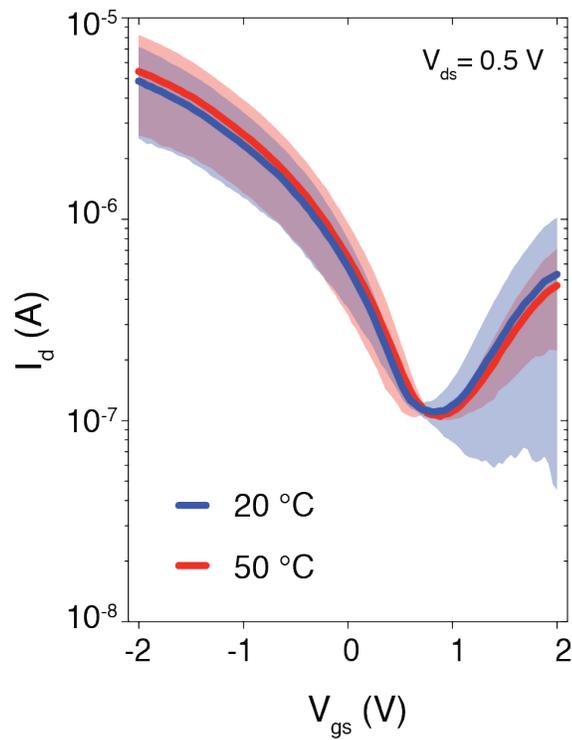

**Fig. 1.**

**Impact of platen temperature while printing all-carbon TFTs**. Subthreshold curves (drain current plotted on a log scale against gate voltage) with transistors printed at 20 °C (blue) and 50 °C (red) platen temperatures, demonstrating no observable difference between the transistor characteristics. Data lines represent average and shaded regions represent standard deviation of 4 devices for both 20 and 50 °C. All devices have a channel length of 250 µm and a channel width of 200 µm.

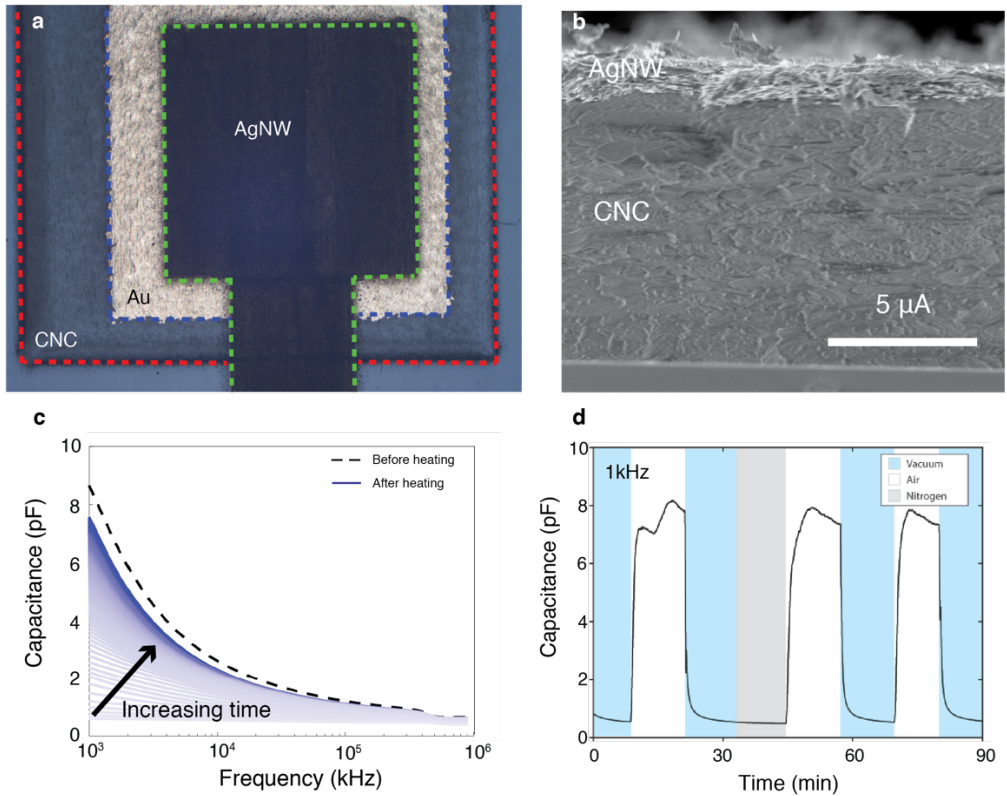

**Fig. 2.**

**CNC dielectric characterization**. (a) Top-view optical image and (b) SEM cross-sectional image of CNC parallel plate capacitor. (c) Capacitance as a function of frequency for CNC capacitor before heating (dashed line) and after heating to 80 °C for 1 hour (solid blue lines) measured every 30 seconds for 30 minutes (light to dark blue), indicating strong frequency dependence of capacitance and a recoverable decay of low-frequency capacitance after heating. (d) Capacitance response to atmosphere conditions in vacuum (blue), air (white), and nitrogen (grey). Both tests demonstrate hydration dependence on capacitance with a decrease in low-frequency capacitance after water evaporation due to heat and vacuum, respectively. The nitrogen atmosphere demonstrates that the capacitance decrease is caused by a hydration decrease as opposed to an environmental pressure change.

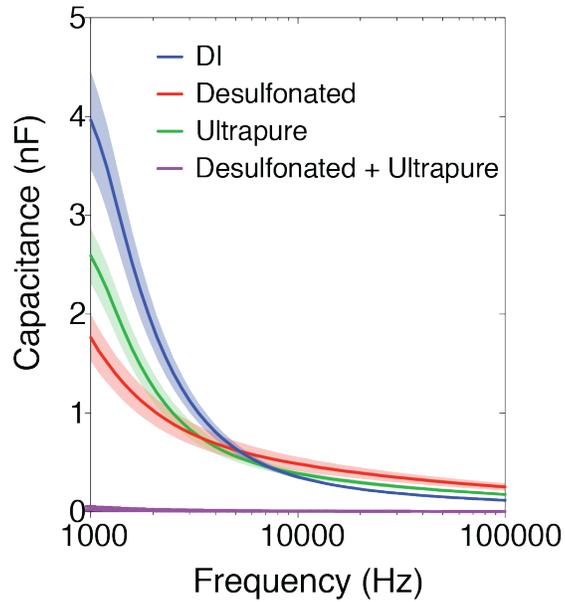

**Fig. 3.**

**CNC capacitance-charge relationship**. CNC capacitance as a function of frequency measured on parallel plate capacitors for CNC dielectric films printed with 6% w/w CNC in DI water (blue), 6% desulfonated CNC in DI water (red), 6% CNC in ultrapure water (green), and 6% desulfonated CNC in ultrapure water (purple). This illustrates that the intrinsic ionic nature of the CNC dielectric is derived from the dissolved salts in addition to the sulfonation from the hydrolysis of cellulose. Data represent average ± standard deviation of 4 devices.

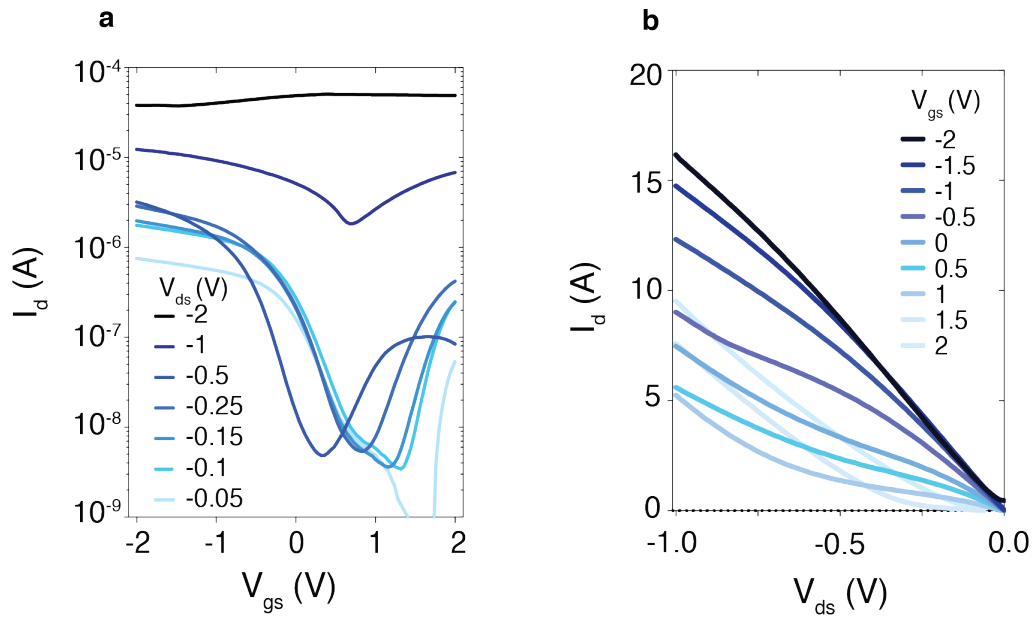

**Fig. 4.**

**CNC-based all-carbon TFT electrical characterization**. (a) Subthreshold curves at varying source-drain voltages ($V_{ds}$) demonstrating the optimization of on/off-current ratio at a $V_{ds}$ of -0.5 V. (b) Output curves at varying gate voltages. A clear shift in conduction pathway can be seen at increasingly positive $V_{gs}$ where carriers are now tunneling through the Schottky barrier into the conduction band rather than the lower-barrier injection of carriers into the valence band at negative gate bias. The transistors were fabricated with a 0.15 mM salt concentration in CNC and tested with a sweep frequency of 10 ms.

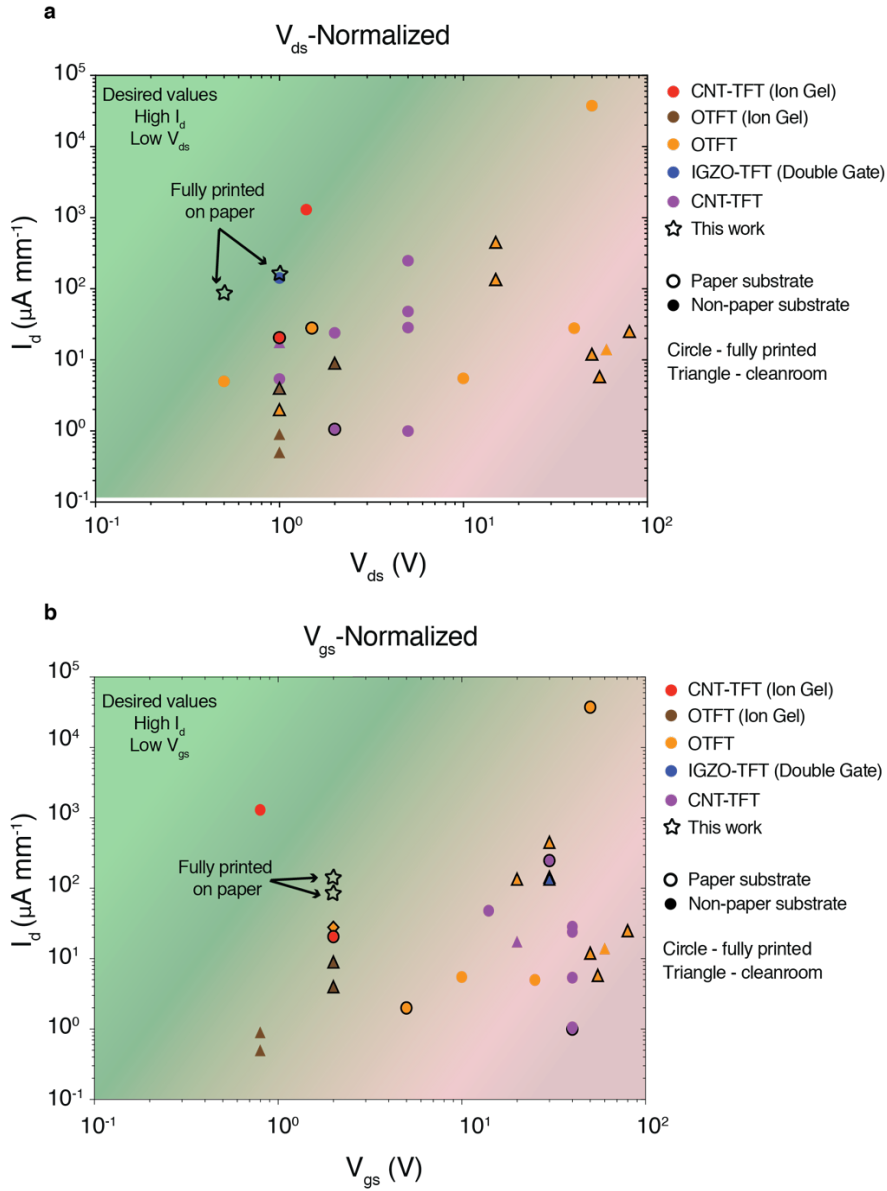

**Fig. 5.**
**Benchmarking printed all-carbon TFT performance against other printed transistors.** (a) Width-normalized on-current plotted against source-drain voltage for related (thin-film transistors) TFTs. (b) Width-normalized on-current plotted against gate voltage. Colors represent transistor types (both semiconducting material and gating scheme); carbon nanotube TFTs (CNT-TFTs) use a semiconducting carbon nanotube channel, organic TFTs (OTFTs) use an organic semiconductor channel and IGZO is indium gallium zinc oxide. Ion gel-based transistors use an ionic liquid as the dielectric. Shapes represent processing technique where circles demarcate fully printed transistors and triangles represent transistors that utilize cleanroom techniques in at least some of the processing. Finally, datapoints with a border are transistors on a paper substrate.

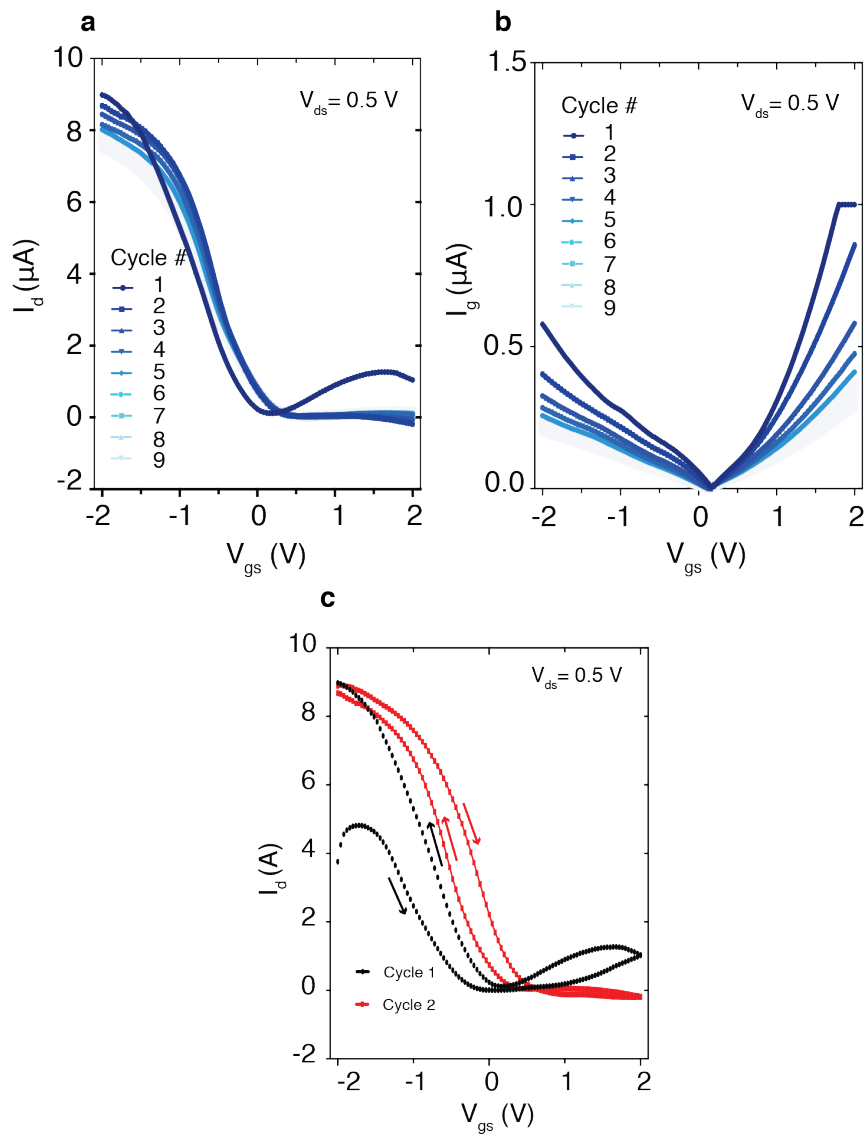

**Fig. 6.**

**CNC-based all-carbon TFT back-to-back cycling**. (a) Transfer curves (drain current plotted on a linear scale against gate voltage) and (b) corresponding gate leakage currents demonstrating a 15% decrease in on-current and a 77% decrease in leakage current over 9 cycles, attributed to limitations in ion mobility and a time delay in the development of a double layer in the CNC dielectric. (c) Transfer curves for two back-to-back cycles showing a marked decrease in hysteresis between the (black) first and (red) second cycles and a significant (~2x) increase in on-current, attributed to the slow formation of the double layer charge that guides the decrease in gate leakage. Without the addition of salt, the transistor behavior is highly dependent on device history, both in terms of cycle duration (hold time) and in terms of cycle number. The transistor was fabricated without added salt and tested with a sweep frequency of 500 ms.

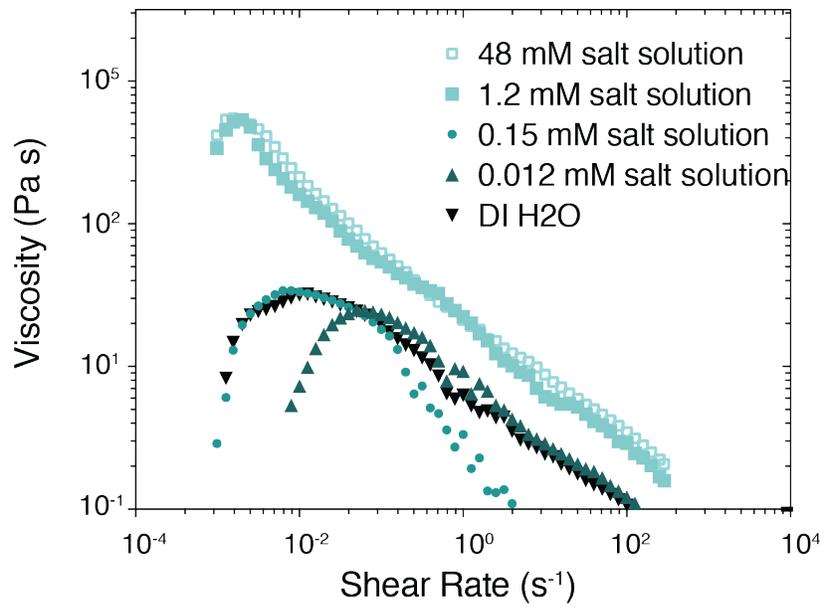

**Fig. 7.**

**Viscosity of CNC ink with added mobile ions**. Dynamic viscosity of 6% CNC solutions with increasing salt concentrations demonstrating the electroviscous effect.

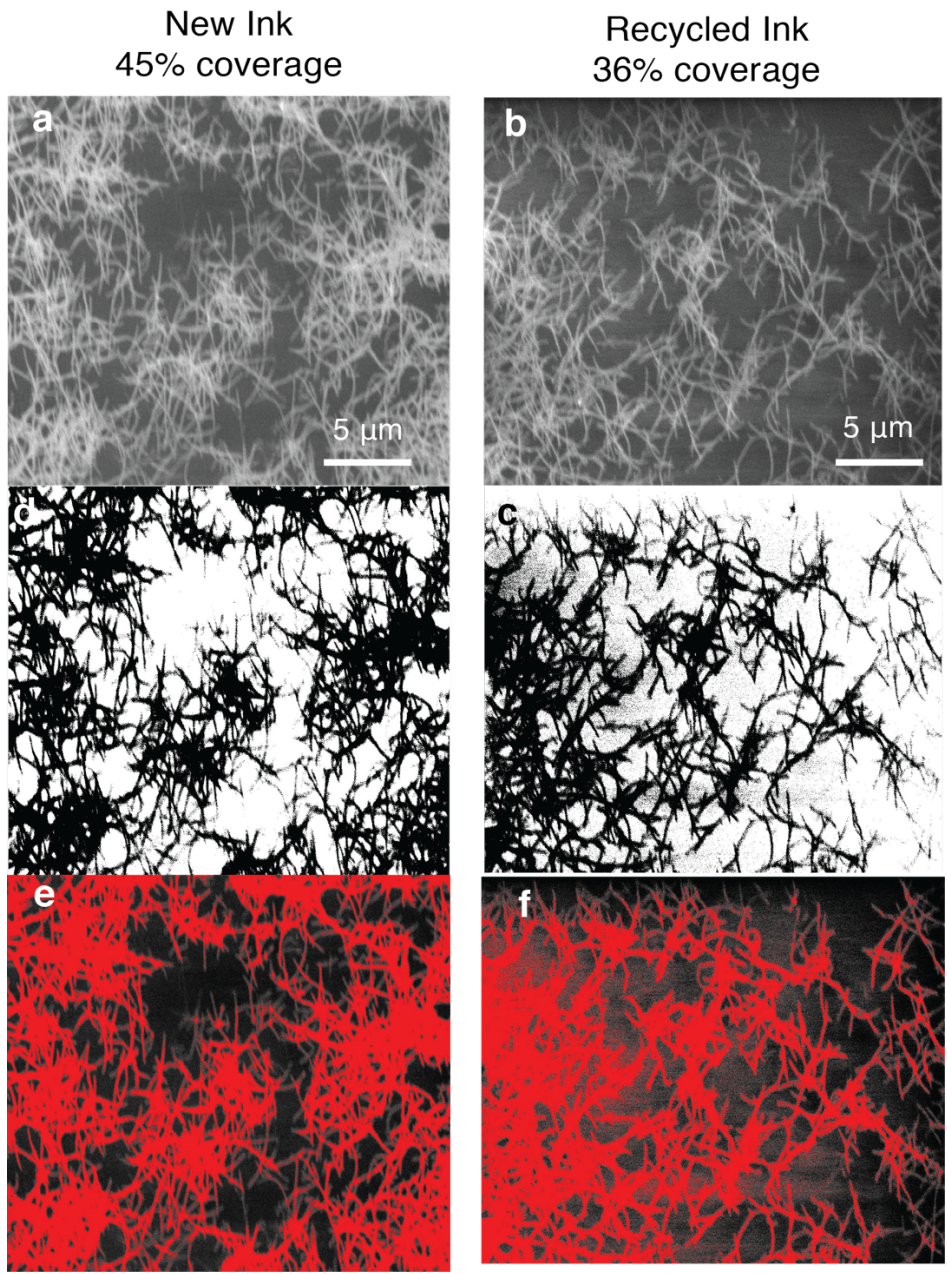

**Fig. 8.**

**CNT printing density.** SEM images of CNT thin films printed on silicon using (a) pristine and (b) recycled CNT ink for purposes of determining printed density (45% and 36% coverage for new and recycled inks, respectively). (c,d) For analysis, each image was set to binary using ImageJ, and in addition, (e,f) the color threshold was modified to help guide the eye. The decrease in printed ink density may help explain the slightly lower on-current with the recycled CNT ink.

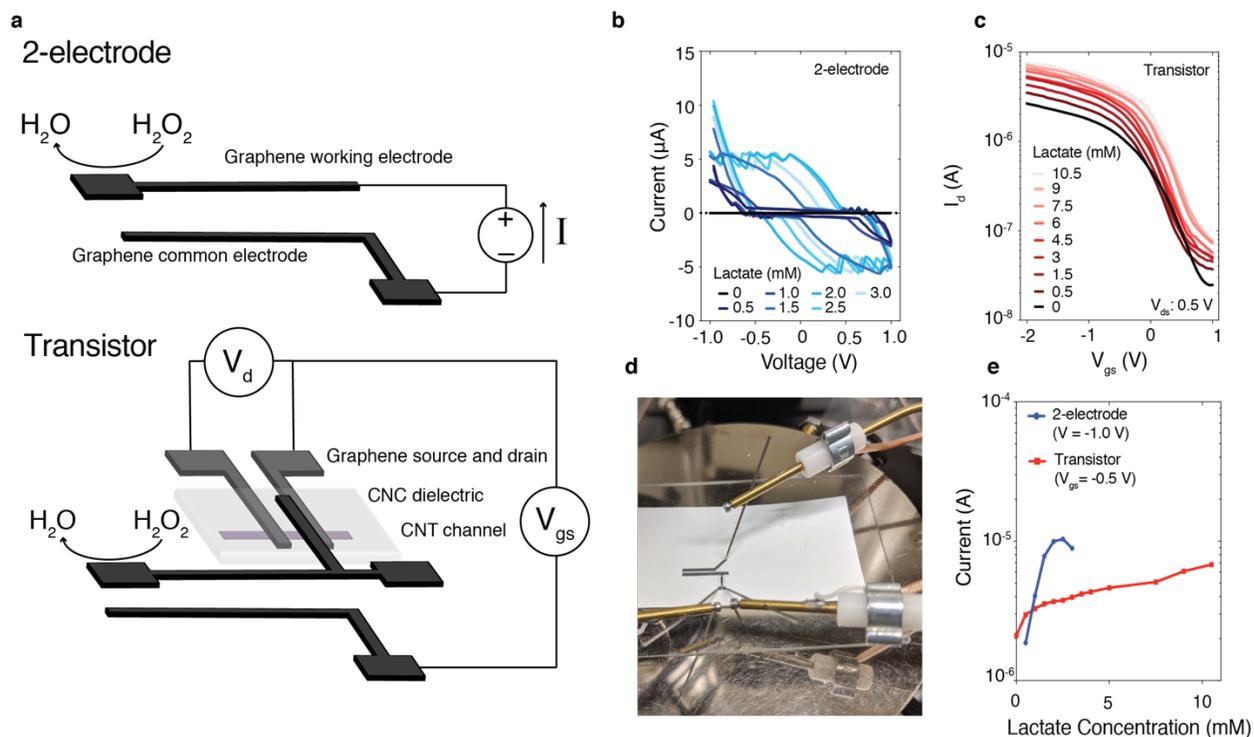

**Fig. 9.**

**Lactate biosensing from fully printed all-carbon devices on paper**. (a) Schematic of (top) 2-electrode and (bottom) transistor-based, all-carbon lactate sensors. Electronic lactate concentration testing using (b) the 2-electrode sensor and (c) the transistor-based sensor. (d) Transistor-based testing setup. e) Current dependence on lactate concentration showing a more pronounced current response to increases in lactate concentration with the 2-electrode device; however, a current maximum is reached at a concentration that is too small for medical relevance whereas the transistor-based device extends the sensing window by at least 5x, allowing for medically relevant lactate sensing. Lactate concentrations greater than 2 mM in human blood are indicative of septic shock *(37)*. Furthermore, recent research indicates that mortality rates for patients presenting with elevated lactate concentrations was highly dependent on the duration of sepsis *(38)*. Mortality sharply declined when lactate levels stabilized to a value of 2.2 mM as opposed to those that maintained a concentration of 3.3 mM after 4 hours. Hence, the 2-electrode sensor has a sensing window that is too narrow for complete medical relevance.

**Extended Data Table 1**.

Comparison table between this work and relevant recent works. (*) marks values that were estimated from graphs in the reports.

| Source | Type of Substrate | Type of Device | Semiconductor | Fabrication procedure | $V_{gs}$ (V) | $V_{ds}$ (V) | On/off ratio | $I_d$ ($\mu A/mm$) | Subthreshold swing (mV/dec) |
|---|---|---|---|---|---|---|---|---|---|
| 1 | Glassine paper | OTFT (Ion Gel) | P3HT | Printed | -2 | -1 | 10^3 | 20.6 | --- |
| 2 | Banknotes (cotton fibre) | OTFT (Ion Gel) | Dinaphtho[2,3-b:2′,3′-f]thieno[3,2-b]thiophene (DNTT) | Cleanroom techniques | 3 | 1 | 10^4 | 4 | 110 mV/dec |
| 3 | Banknotes (cotton fibre) | OTFT (Ion Gel) | dinaphtho[2,3-b:2′,3′-f]thieno[3,2-b]thiophene (DNTT) | Cleanroom techniques | -3 | -2 | 10^5 | 9 | 103 |
| 4 | Photo paper | OTFT | Pentacene | Cleanroom techniques | -55 | -55 | 10^5 | 5.8 | >1000* |
| 5 | Printer paper | OTFT | Pentacene and DNTT | Cleanroom techniques and printing | -80 | -80 | 10^8 | 25 | 1400 |
| 6 | Printer paper | OTFT | pBTT | Printed | -50 | -50 | 10^4 | 37,500 | 18100 |
| 7 | Paper | OTFT | Pentacene | Cleanroom techniques | -5 | 1 | 10^2 | 2 | 600 |
| 8 | PET – CNC dielectric | OTFT | PCDTBT | Cleanroom techniques | -50 | -50 | 10^4 | 12 | --- |
| 9 | Paper | OTFT | GIZO | Cleanroom techniques | 20 | 15 | 10^4 | 135 | 800 |
| 10 | Paper | OTFT | Poly(3,3000-didode- cyl-quarterthiophene | Printed | -2 | -1.5 | 15 | 28 | >1600* |
| 11 | CNC paper | OTFT | GIZO | Cleanroom techniques | 30 | 15 | 10^5 | 450 | 2110 |
| 11 | CNC paper | OTFT | GIZO | Cleanroom techniques | 30 | 15 | 10^4 | 135 | 1790 |
| 12 | Eucalyptus paper | TFT (double gate) | IGZO | Cleanroom techniques | 30 (15V back gate) | 1 | 10^5 | 145 | 1600 |
| 13 | Photo paper | CNT-TFT | CNT | Printed | -40 | -2 | 10^5 | 1 | ~9000* |
| 14 | CNTs embedded into cellulose | CNT-TFT | CNT | Solution processed | -5 | 9 | 10^2 | --- | --- |
| **This work** | **Photo paper** | **Ion-gel** | **CNT** | **Printed** | **2** | **0.5** | **10^3** | **87** | **138** |
| 15 | Si | Ion-gel | CNT | Printed | -0.8 | -1.4 | 10^5 | 1300 | <150 |
| 16 | Polyimide | OTFT (Ion Gel) | DPh-BTBT | Printed and cleanroom. techniques | 1 | -1 | 10^6 | 0.9 | 90 |
| 16 | Polyimide | OTFT (Ion Gel) | N1100 | Printed and cleanroom. techniques | 1 | 1 | 10^6 | 0.5 | 100 |
| 17 | Si | OTFT | CNT | Printed | -25 | -0.5 | 10^3 | 5 | ~5000* |
| 18 | Paralyene-C | OTFT | Merck lisicon S1200 | Printed | -10 | -10 | 10^6 | 5.5 | ~750* |
| 19 | Polycarbonate | OTFT | TIPS-Pentacene | Printed | -60 | -60 | 10^5 | --- | >1500* |
| 20 | PEN | OTFT | P(NDI2OD-T2) | Printed and cleanroom. techniques | 60 | 60 | 10^4* | 14 | --- |
| 21 | PMMA | OTFT | C8-BTBT | Solution processed | -40 | -40 | 10^4 | 28 | >1000* |
| 22 | Polyimide | CNT-TFT | CNT | Printed | 40 | 5 | 10^4 | 28.6 | ~750* |
| 23 | Polyimide | CNT-TFT | CNT | Printed | -30 | -5 | 10^6 | 248 | --- |
| 24 | Si | CNT-TFT | CNT | Printed | -40 | -1 | 10^4 | 5.4 | ~6000* |
| 25 | Polyimide | CNT-TFT | CNT | Printed | -40 | -5 | 10^4 | 1 | ~3000* |
| 26 | Si | CNT-TFT | CNT | Printed | -40 | -2 | 10^5 | 24 | ~4000* |
| 27 | PET | CNT-TFT | CNT | Printed | -14 | -5 | 10^4 | 48 | >1000* |
| 28 | Si | CNT-TFT | CNT | Printed and cleanroom. techniques | -20 | 1 | 10^5 | 17.6 | ---- |